\documentclass[times, twoside, watermark]{zHenriquesLab-StyleBioRxiv}

\leadauthor{Ning and Jorfi} 

\begin{document}

\title{p75NTR as a Molecular Memory Switch}
\shorttitle{Memory Switch}

\author[1,2]{Shen Ning}
\author[1,3,\Letter]{Mehdi Jorfi}

\affil[1]{Department of Neurology, Massachusetts General Hospital, Harvard Medical School, Charlestown, Massachusetts, United States}
\affil[2]{Boston University School of Medicine, Boston, Massachusetts, United States}
\affil[3]{Center for Engineering in Medicine, Massachusetts General Hospital, Harvard Medical School, Charlestown, Massachusetts, United States}

\maketitle

\begin{abstract}
In recent years, many molecular and environmental factors have been studied to understand how synaptic plasticity is modulated. Sleep, as an evolutionary conserved biological function, has shown to be a critical player for the consolidation and filtering of synaptic circuitry underlying memory traces. Although sleep disturbances do not alter normal memory consolidation, they may reflect fundamental circuit malfunctions that can play a significant role in exacerbating diseases, such as autism and Alzheimer’s disease. Very recently, scientists sought to answer part of this enigma and they identified p75 neurotrophic receptor (p75NTR) as a critical player in mediating impairments in hippocampal-dependent associative plasticity upon sleep deprivation. This paper will review the role of the p75NTR, critically discuss the impact and implications of this research as the bridge for sleep research and neurological diseases. 
\end {abstract}


\begin{corrauthor}
mjorfi\at mgh.harvard.edu
\end{corrauthor}

\section*{Introduction}
Synaptic plasticity, defined as experience-dependent changes in neuronal connectivity, is fundamental to learning and memory. The hippocampus is a brain structure in the medial temporal lobe of the brain that has an essential role in memory formation. Previous studies have established three well-accepted synaptic pathways, including the performant, mossy fiber, and Schaffer collateral pathways \cite{Frey}. These pathways have distinct layers and cellular projections that enable synaptic strengthening through long-term potentiation (LTP) or synaptic weakening via long-term depression (LTD) \cite{Ho}. There are two generally accepted phases of LTP: early-LTP, which lasts less than a few hours and depends primarily on short-term kinase activity and late-LTP, which lasts more than eight hours after the inducing stimulus and depends on the activation of gene transcription \cite{Huang}. The persistence of memory is reliant on the synaptic potentiation that triggers intracellular signal transduction cascades during late-LTP. The synaptic tagging and capture (STC) hypothesis assert that LTP occurs due to the sequential expression of synaptic potentiation with a synaptic tag. This is followed by the synthesis and distribution of plasticity-related proteins and the capture of these proteins by the tagged synapses, leading to the preservation of the synaptic strength \cite{Redondo}.
In recent years, many molecular and environmental factors have been studied to understand how synaptic plasticity is modulated. Sleep, as an evolutionary conserved biological function, has shown to be a critical player for the consolidation and filtering of synaptic circuitry underlying memory traces. Although sleep disturbances do not alter normal memory consolidation, they may reflect fundamental circuit malfunctions that can play a significant role in exacerbating diseases, such as autism and Alzheimer’s disease \cite{Ning:20199c3}. Sleep researchers have found that non-rapid eye movement (NREM) sleep, which is characterized by slow waves of electrical activity throughout the cortex, increases hippocampal firing, while rapid eye movement (REM) sleep, which is characterized by rapid eye movement, muscle paralysis, and low voltage irregular brain waves, decrease firing rate in the hippocampus \cite{Wang:2011cn}. Sleep disruption, particularly of REM sleep, has been shown to impair late-LTP, suggesting that diseases that disrupt a specific sleep cycle may have more detrimental effects \cite{Gronli} .

\section*{p75NTR is a Key Molecular Signal for Synaptic Plasticity During Sleep Deprivation}
While previous research has robustly established the connection between sleep and memory consolidation, little is known about the molecular mechanisms underlying this process. A recent paper published by Wong and colleagues \cite{Wong} sought to answer part of this enigma. They identified p75 neurotrophic receptor (p75NTR) as a critical player in mediating impairments in hippocampal-dependent associative plasticity upon sleep deprivation. p75 is a neurotrophic receptor responsible for mediating neural survival and apoptosis through the neurotrophin Trk receptor pathway \cite{Bernabeu}. p75NTR is expressed by dentate progenitor cells and is an essential mediator for proliferation and maturation of neural, glial, and other cell types \cite{Bernabeu}. It also serves as a co-receptor of both the sortilin for proneurotrophins to mediate apoptosis and the Nogo receptor to mediate neurite collapse \cite{Dechant}. Interestingly, p75NTR is re-expressed in adults during pathological states such as epilepsy, axotomy, and neurodegenerative diseases, indicating its crucial role in mediating neuronal death, axonal degeneration and dysfunction, synaptic transmission and plasticity modulation via LTP \cite{Dechant}. Specifically, p75NTR binds with amyloid-beta (Aβ) in Alzheimer’s disease and regulates Aβ-mediated degeneration of cholinergic neurons \cite{Yao}. While a previous work established that circadian rhythm can regulate the transcriptional activation of p75NTR in the nervous system and peripheral tissues, indicating its role in maintenance of the circadian clock and metabolic gene oscillation, the association between p75NTR and sleep deprivation in synaptic plasticity and cognitive decline was not investigated \cite{Baeza-Raja}. In their paper, Wong et al. \cite{Wong} showed that sleep deprivation could impair crucial synaptic processes that are necessary for associative learning and memory. Furthermore, they found that sleep deprivation modulates the RhoA-ROCK2 and cAMP-PKA-LIMK1-cofilin pathways via p75NTR.

\section*{p75NTR Confers a Protective Effect on Late LTP}
Sleep deprivation impairs memory consolidation by inhibiting the synthesis of proteins necessary for synaptic plasticity, thereby reducing neuronal excitability and synaptic strength formation \cite{McDermott, Abel}. Notably, previous studies investigated the molecular underpinnings and found that sleep deprivation selectively impaired cAMP/PKA-dependent forms of synaptic plasticity in the hippocampus to reduce cAMP signaling and increase phosphodiesterase-4 activity, which degrades cAMP \cite{Vecsey}. One key finding in the recently published paper by Wong et al. \cite{Wong} is that p75NTR confers a protective effect on late LTP deficits induced by sleep deprivation. Sleep deficiency can cause LTP deficits, specifically STC deficits. Interestingly, the authors showed that the deletion of p75NTR in mice prevents late LTP and STC deficits induced by sleep shortage. The mechanism proposed in this study complements previous work on sleep deprivation and highlights a vital aspect of the specific mechanisms of LTP that is interrupted by sleep deprivation. Not only does this study answer a fundamental question about the function of sleep for memory consolidation, but it is also informative regarding how neurological diseases that disrupt sleep can lead to cognitive decline. However, sleep deficits are commonly found in many neurological diseases, suggesting that it may play an indirect causative role in promoting and exacerbating diseases. Alternatively, it can be argued that diseases may have different causes that lead to sleep disruption and serve as a critical contributing factor to cognitive deficits, particularly in memory.

\section*{p75NTR Deletion Restores Associative Memory}
Wong and colleagues \cite{Wong} also assessed the behavioral changes in mice with p75NTR deletion and noted that p75NTR deletion could restore associative memory in animals suffering from sleep deprivation. To achieve this, they used a behavioral tagging paradigm where mice are exposed to a different environment or open field before application of a foot shock. To examine the associated memory, they used the latency to step down to bars since prior studies have shown that a weak memory, such as a weak foot shock, can become a stronger memory is coupled to a strong stimulus, such as an open field. The longer the step-down latency, the stronger the associative memory. Typically, sleep deprivation reduces step-down latency, but in p75NTR knockout mice, the authors did not observe this reduction, suggesting that the mice retained normal memory. While the evidence is convincing, the methods of the study are limited to demonstrating associative memory in a very basic paradigm. It will be crucial to investigate further whether sleep deprivation selectively hinders associative memory specifically or if it affects all memory types. If it contributes to other cognitive changes, future work should examine whether knocking out p75NTR can also rescue other cognitive functions. Another critical question that needs to be addressed in future studies is the effect of knocking out p75NTR on sleep staging, mainly because p75NTR is expressed in the brainstem, which regulates sleep staging and arousal. Finally, it would be interesting to isolate REM and NREM disruption and observe its molecular effects on synaptic plasticity and determine whether if knocking out p75NTR can rescue these effects. 

\section*{p75NTR Deletion Prevents Reduction of Hippocampal Spine Density}
Sleep deprivation increases phosphodiesterase 4A5 (PDE4A5) in the hippocampus to decrease cAMP levels, which binds to the cAMP-responsive element-bind protein (CREB). These proteins mediate synaptic plasticity and memory \cite{Bramham}. CREB is a transcription factor that has numerous regulatory effects, including regulating brain-derived neurotrophic factor (BDNF), a well-known factor that regulates the induction as well as early and late maintenance phases of LTP expression \cite{Bramham}. Previous studies have found that knocking out p75NTR in mice impaired NMDA receptor-dependent long-term depression (LTD) through selective downregulating NR2B, an NMDA receptor subunit uniquely involved in LTD. More interestingly, proBDNF activates p75NTR to enhance NR2B-dependent LTD and NR2B-mediated synaptic currents \cite{Woo}. Wong et al. \cite{Wong} found that the deletion of p75NTR prevents SD-mediated changes in PDE4A5-cAMP-CREB and BDNF pathway. Furthermore, Wong and colleagues showed that p75NTR deletion could prevent the reduction of hippocampal spine density by reversing the harmful modulation of the RhoA-ROCK2 and cAMFP-PKA-LIMK1-cofilin pathways that are important for actin dynamics \cite{Wong}. This molecular connection is particularly relevant for neurological diseases such as Alzheimer’s disease, which exhibits decreased disassembly of f-actin to contribute to cognitive decline \cite{Kommaddi}. However, sleep disruption is a universal symptom for a variety of diseases, which suggests that sleep deprivation is a secondary symptom that can exacerbate cognitive decline even though it may not be necessarily the root cause of it. Figure 1 shows the association between sleep deprivation, LTP, and associative learning and memory in wildtype mouse and knockout mouse lacking p75NTR as well as the breakdown of the molecular pathways post-sleep deprivation.

\begin{figure*}
\centering
\includegraphics[width=1.0\linewidth]{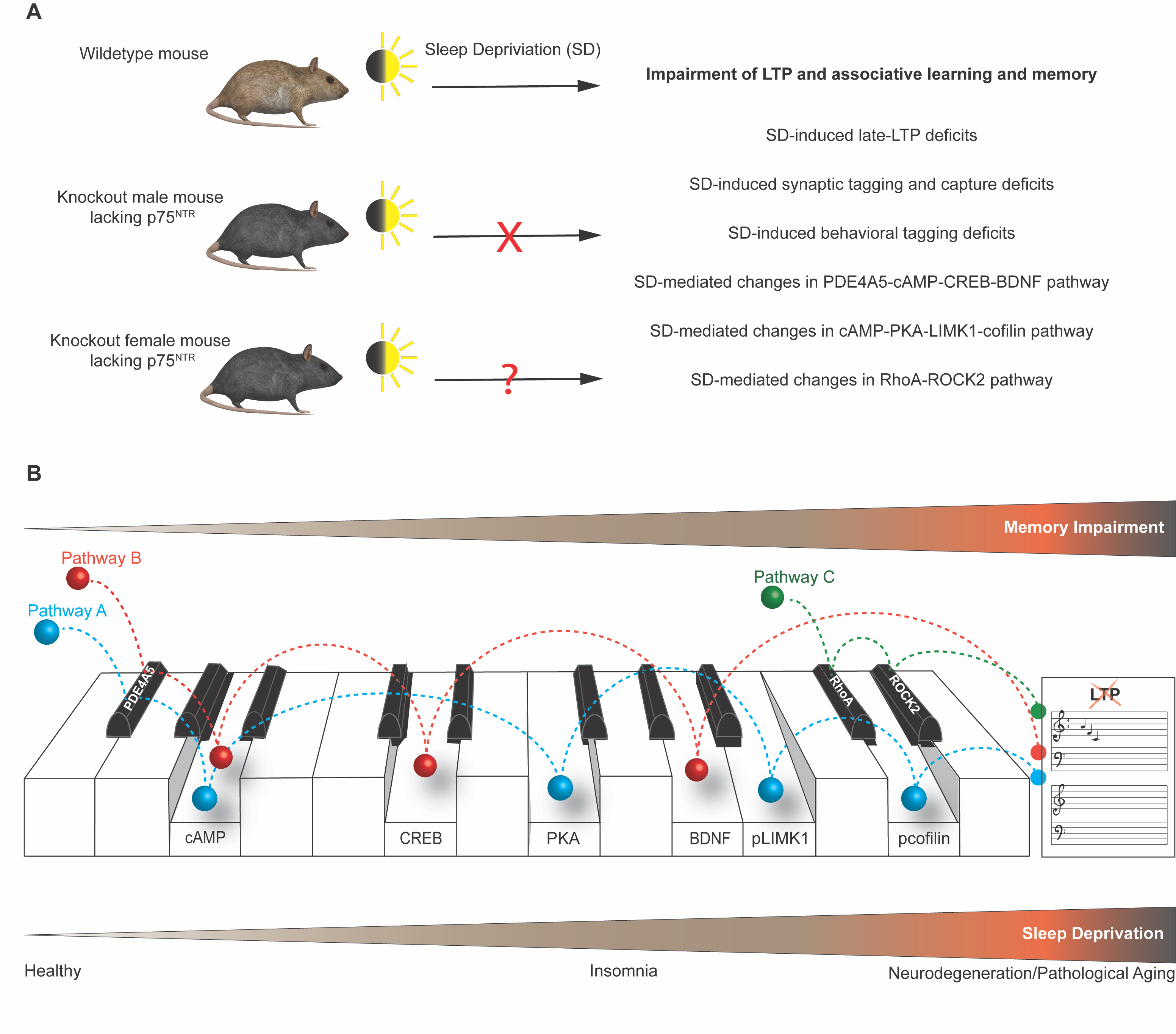}
\caption{Impairment of LTP and associative learning and memory. A, Sleep deprivation induces late LTP deficits and synaptic tagging and captures deficits. Sleep deprivation also alters the PDE4A5-cAMP-CREB-BDNF, cAMP-PKA-LIMK1-cofilin, and RhoA-ROCK2 pathways. This Figure, in part, used Servier Medical Art, which is licensed under a Creative Commons Attribution 3.0 Unported License: https://smart.servier.com B, Breakdown of the molecular pathways post-sleep deprivation. PDE4A5 is upregulated, leading to decreased cAMP followed by decreased CREB, and finally downregulation of BDNF (Pathway A: PDE4A5-cAMP-CREB-BDNF). In parallel, decreased cAMP leads to decreased PKA, downregulating pLIMK1, and then cofilin (Pathway B: cAMP-PKA-LIMK1-cofilin). Pathway C shows that sleep deprivation causes the upregulation of RhoA, which results in increased ROCK2 (Pathway C: RhoA-ROCK2). All these three pathways lead to impaired LTP and associative memory deficits.}
\label{fig:computerNo}
\end{figure*}

\section*{Current Limitations}
In the latest study by Wong et al. \cite{Wong}, the authors used the exon III p75 KO mouse model, which fails to show some crucial signaling activity and display phenotypical characteristics that confound the study. Other mutant animal models, such as the conditional p75 mice which carry a conditional p75NTR (p75NTR-FX/FX) allele made by flanking exons 4-6 instead of 3, show a complete deletion of the transmembrane and cytoplasmic domains of the receptor without the molecular complexities observed in mice with the p75NTRΔexon3 allele \cite{Bogenmann}. It would be interesting to compare the impact of sleep deprivation in this conditional knockout model with the model used in this study to explore the specificity of the rescue effect reported in this paper. Future studies should also examine the chronic effects of sleep deprivation on these molecular pathways and determine if they propagate further downstream to pathways that are crucial for healthy brain function. This can be compared with the acute effects to determine the chronic and acute physiological roles of p75NTR on synaptic plasticity.

\section*{Conclusions and Future Directions}
The crucial role of p75NTR in mediating memory deficits makes this marker a potential future therapeutic target for neurological diseases that disrupt normal sleep patterns. However, it remains unknown what the off-target effects and whether long-term treatment for chronic diseases would have an overall beneficial impact. It is also unclear whether the results of this study will be translatable in a human clinical setting. The physiological relevance of the mouse model used in this study was not addressed. Indeed, we can bring in genetic modification using delivered CRISPR or other gene silencing techniques. However, the therapeutic relevance of these molecules is still in their infancy, and scientists and policymakers alike should critically evaluate the ethical implications of these forms of biotechnology. Another nuanced indication is that many neurological diseases are both diseases of molecular and protein deficits and changes to the circuitry. Sleep deficiency seems to be a crucial link bridging molecular alteration with circuit function for memory consolidation. Future experiments can use new tools and task paradigms, such as optogenetics, to explore whether this effect also pertains to spatial and contextual memory, and how engram-bearing cells are affected, as well as whether p75NTR is essential for both the establishment of the initial memory trace and the consolidation of the memory. Once established, these future findings may inform us about the translational capability of this molecule to prevent memory loss and cognitive decline typical of aging and numerous devastating neurological diseases and disorders.

\section*{Disclosures}
No conflicts of interest, financial or otherwise, are declared by the authors.

\section*{Authors Contributions}
Both authors contributed to the design of the manuscript including acquisition of references and figure. S.N. and M.J. contributed to drafting and critically revising the manuscript and approved the final version. 

\section*{References}
\bibliography{mybibliography.bib}


\end{document}